\documentclass[conference]{IEEEtran}
\pdfoutput=1
\usepackage{cite}
\usepackage{amsmath,amssymb,amsfonts}
\usepackage{algorithmic}
\usepackage{graphicx}
\usepackage{textcomp}
\usepackage{xcolor}
\usepackage{booktabs}
\usepackage{tabularx}
\usepackage{array}
\usepackage{tikz}
\usetikzlibrary{shapes.geometric,arrows.meta,positioning,calc,fit,backgrounds}
\newcommand{\figtiny}{\fontsize{6.0pt}{6.8pt}\selectfont}
\newcommand{\para}[1]{\smallskip\noindent\textit{#1}\hspace{0.4em}}
\def\BibTeX{{\rm B\kern-.05em{\sc i\kern-.025em b}\kern-.08em
    T\kern-.1667em\lower.7ex\hbox{E}\kern-.125emX}}
\begin{document}

\title{Securing and Auditing Long-Horizon LLM Agents:\\ State-Transfer Auditing and Controls in the Agent Harness}

\author{\IEEEauthorblockN{Deep Shah}
\IEEEauthorblockA{\textit{Google LLC}\\
Mountain View, CA, United States \\
shahdeep@google.com}
\and
\IEEEauthorblockN{Nehal Kathrotia}
\IEEEauthorblockA{\textit{Google LLC}\\
Mountain View, CA, United States \\
nehalk@google.com}
}

\maketitle

\begin{abstract}
Long-horizon LLM agents across codebases, web browsers, and enterprise infrastructure rely on an execution harness to move state between the working context, session history, persistent memory, and external environment. While conventional execution traces log model prompts and tool calls, critical security and reliability failures occur during state transfers: a transfer can admit unchecked content (indirect prompt injections, unverified memory writes), rewrite state lossily (context compaction dropping user constraints, memory consolidation stripping source authority), or restore state incompletely (workspace rollbacks leaving aborted KV state active on the inference server). Because current traces omit what crosses these boundaries, failure attribution and runtime monitors fail on multi-step trajectories. We present a state-transfer audit architecture for the agent harness. At each transfer boundary, we define an audit point specifying the minimal hash-chained record the harness emits, separating low-overhead deterministic checks from model-based semantic checks and binding internal context, memory, and inference-server cache state to external actions. We show that the same transfer records required for causal failure attribution supply the durable state needed by runtime security controls, including least-privilege capabilities, replay-resistant authorization outside the prompt, information flow control, obligation monitors, and transactional recovery across multi-agent and infrastructure pipelines. Finally, we analyze multimodal GUI adaptations and implement and evaluate a reference harness middleware across 240 multi-step fault-injection trajectories, showing that transfer records localize root causes and catch cross-transfer violations missed by lifecycle and tool-only logs at microsecond deterministic overhead.
\end{abstract}


%=====================================================================
\section{Introduction}\label{sec:intro}

LLM agents now resolve repository issues, operate websites and desktop applications, and automate IT operations, often over many steps. Each tool and data source they read widens the attack surface~\cite{arx2502_08586}: AIOps agents, for example, can be steered through manipulated telemetry~\cite{arx2508_06394}. Task completion rates do not show whether an agent behaved safely without unauthorized side effects~\cite{arx2610_13623}. An operator must also establish, after the fact, why an action was taken, which input caused it, and whether a recovery restored a safe state. Answering these questions depends on what the code around the model records.

We use the term \emph{harness} for all code that executes between model calls: it parses tool calls, runs them, formats observations, assembles the prompt, compacts history, and reads and writes memory. The harness shapes both cost and accuracy: holding the model fixed, coding harnesses differed by up to 40$\times$ in tokens per solved task~\cite{scaffold2026effect}, and model rankings reversed across harnesses~\cite{arx2610_00917}. It is also where audit evidence is produced or lost.

Current traces support attribution poorly. On the Who\&When benchmark of multi-agent failure logs, the best of three initial methods identified the failure-responsible step in only 14.2\% of cases~\cite{zhang2025which}, and long-context models performed poorly at locating errors in the human-annotated traces of TRAIL~\cite{arx2505_08638}. The traces themselves omit key fields. When existing agent trajectories were replayed through a temporal-logic monitor, the logs rarely included approvals and never included timestamps, reducing history-dependent policies to flagging risky action types~\cite{arx2610_09793}. Existing observability schemas trace events across the agent lifecycle (Section~\ref{sec:related}), without tying each record to a failure it detects.

This paper presents a state-transfer audit architecture across four harness stores:
\begin{itemize}
\item a state-transfer failure model across coding and GUI harnesses covering three failure modes: admitting unchecked content, lossy rewriting, and incomplete restoration (Sections~\ref{sec:transfers}--\ref{sec:fail}, Table~\ref{tab:evidence});
\item transfer-level \emph{audit points} specifying emitted hash-chained record fields, deterministic versus model-based checks, and enforced controls across all eight transfers (Section~\ref{sec:audit}, Table~\ref{tab:audit});
\item how transfer records directly support causal failure attribution and runtime obligation monitoring (Section~\ref{sec:obs});
\item how the same records supply durable runtime state for least-privilege capabilities, replay-resistant authorization, information flow control, tamper-evident accountability, and recovery across multi-agent and infrastructure pipelines (Section~\ref{sec:security}, Table~\ref{tab:props});
\item multimodal GUI adaptations (Section~\ref{sec:mm}), a wrapper-based implementation architecture (Section~\ref{sec:impl}), and an empirical evaluation of wrapper overhead, violation detection, and attribution ablations against standard tracing schemas (Section~\ref{sec:eval}, Tables~\ref{tab:overhead}--\ref{tab:eval}).
\end{itemize}

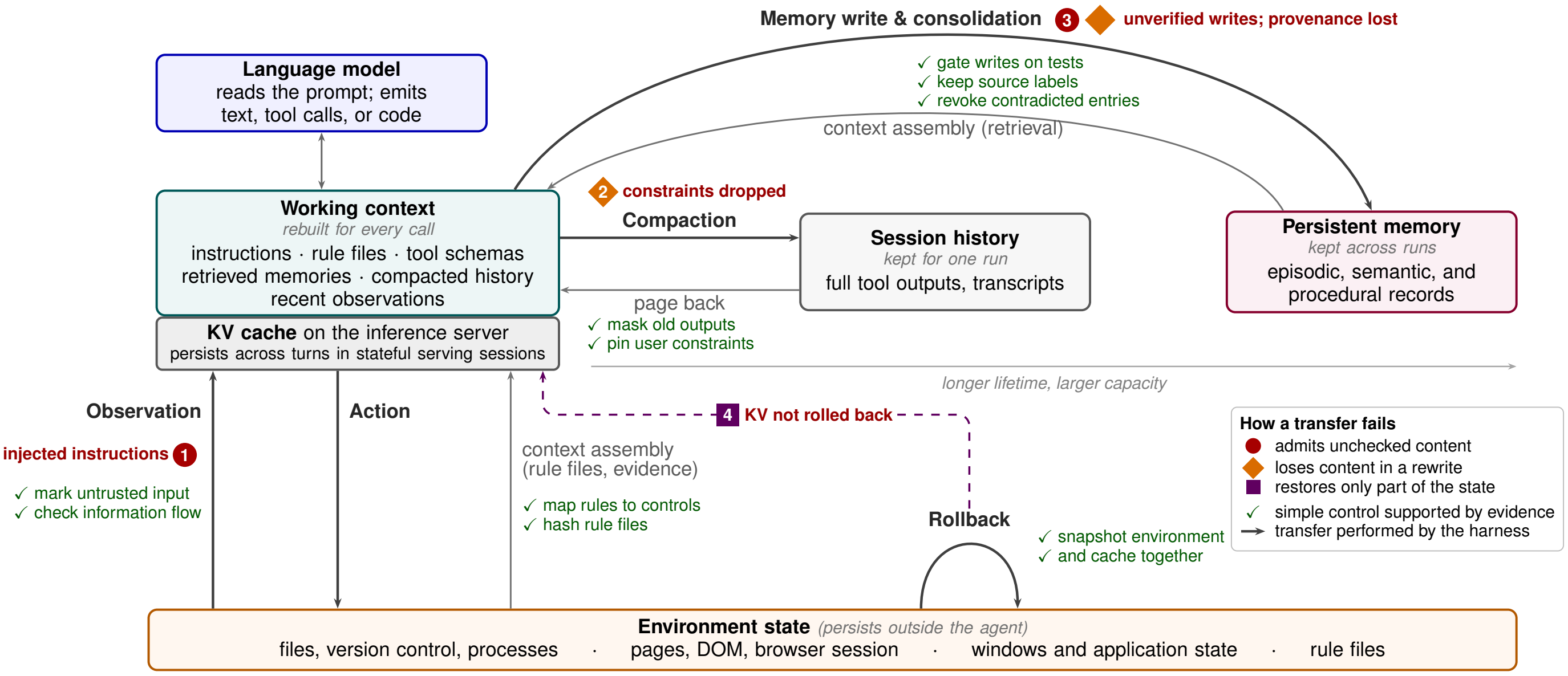
\begin{figure*}[!t]
\centering
\resizebox{\textwidth}{!}{%
\begin{tikzpicture}[
    font=\sffamily\scriptsize,
    place/.style={align=center, rectangle, rounded corners=3pt, line width=0.8pt, inner sep=2.5pt},
    flow/.style={draw=black!75, -{Stealth[length=3.6pt]}, line width=0.9pt},
    thin/.style={draw=black!55, -{Stealth[length=3.2pt]}, line width=0.6pt},
    lab/.style={align=center, font=\sffamily\scriptsize\bfseries, text=black!85, inner sep=1pt},
    sub/.style={align=center, font=\sffamily\scriptsize, text=black!65, inner sep=1pt},
    ok/.style={font=\sffamily\figtiny, text=green!35!black, inner sep=1pt, align=left},
    bad/.style={font=\sffamily\figtiny\bfseries, text=red!60!black, inner sep=1pt, align=left},
    admit/.style={circle, fill=red!65!black, text=white, font=\sffamily\figtiny\bfseries, inner sep=0pt, minimum size=8.5pt},
    lossy/.style={diamond, fill=orange!85!black, text=white, font=\sffamily\figtiny\bfseries, inner sep=0pt, minimum size=11pt},
    incomplete/.style={rectangle, fill=violet!75!black, text=white, font=\sffamily\figtiny\bfseries, inner sep=0pt, minimum size=8pt},
]

% ================= Places of state =================
\node[place, draw=blue!70!black, fill=blue!6, minimum width=4.1cm, text width=3.9cm, minimum height=0.8cm] (model) at (2.55,0.35)
  {\textbf{Language model}\\reads the prompt; emits text, tool calls, or code};

\node[place, draw=teal!70!black, fill=teal!7, minimum width=5.0cm, text width=4.8cm, minimum height=1.55cm, anchor=north] (wc) at (3.0,-0.85)
  {\textbf{Working context}\\[-1pt]{\figtiny\itshape\color{black!55}rebuilt for every call}\\[1pt]
   instructions $\cdot$ rule files $\cdot$ tool schemas\\ retrieved memories $\cdot$ compacted history\\ recent observations};
\node[place, draw=gray!70!black, fill=gray!13, minimum width=5.0cm, text width=4.8cm, minimum height=0.55cm, anchor=north, rounded corners=2pt] (kv) at (wc.south)
  {\textbf{KV cache} on the inference server\\[-1pt]{\figtiny persists across turns in stateful serving sessions}};

\node[place, draw=gray!70!black, fill=gray!6, minimum width=3.6cm, text width=3.4cm, minimum height=1.2cm] (sess) at (10.3,-1.75)
  {\textbf{Session history}\\[-1pt]{\figtiny\itshape\color{black!55}kept for one run}\\[1pt]full tool outputs, transcripts};
\node[place, draw=purple!70!black, fill=purple!6, minimum width=3.6cm, text width=3.4cm, minimum height=1.2cm] (mem) at (15.6,-1.75)
  {\textbf{Persistent memory}\\[-1pt]{\figtiny\itshape\color{black!55}kept across runs}\\[1pt]episodic, semantic, and\\procedural records};

\node[place, draw=orange!70!black, fill=orange!6, minimum width=17.0cm, text width=16.8cm, minimum height=0.75cm] (env) at (8.9,-6.45)
  {\textbf{Environment state} {\figtiny\itshape\color{black!55}(persists outside the agent)}\\
   files, version control, processes $\;\cdot\;$ pages, DOM, browser session $\;\cdot\;$ windows and application state $\;\cdot\;$ rule files};

% lifetime axis
\draw[draw=black!35, -{Stealth[length=3pt]}, line width=0.5pt] (5.9,-3.05) -- (17.4,-3.05)
  node[pos=0.5, below=0pt, font=\sffamily\figtiny\itshape, text=black!50] {longer lifetime, larger capacity};

% ================= Model <-> working context =================
\draw[draw=black!60, {Stealth[length=3.2pt]}-{Stealth[length=3.2pt]}, line width=0.6pt] (model.south) -- (model.south |- wc.north);

% ================= Compaction / page back =================
\draw[flow] (wc.east |- 0,-1.45) -- (sess.west |- 0,-1.45) node[midway, above=1pt, lab] {Compaction};
\node[lossy] (m2) at (6.05,-0.88) {2};
\node[bad, anchor=west] at (m2.east) {constraints dropped};
\draw[thin] (sess.west |- 0,-2.1) -- (wc.east |- 0,-2.1) node[midway, below=0.5pt, sub] {page back};
\node[ok, anchor=north west] at (5.8,-2.4) {$\checkmark$ mask old outputs\\$\checkmark$ pin user constraints};

% ================= Memory write (outer arc) / retrieval (inner arc) =================
\draw[flow] (4.95,0 |- wc.north) .. controls (6.6,1.75) and (14.4,1.75) .. (15.6,0 |- mem.north)
  node[pos=0.45, above=1pt, lab] (mwl) {Memory write \& consolidation};
\node[admit, anchor=west] (m3) at ([xshift=3pt]mwl.east) {3};
\node[lossy, anchor=west] (m3b) at ([xshift=1.5pt]m3.east) {};
\node[bad, anchor=west] at ([xshift=1.5pt]m3b.east) {unverified writes; provenance lost};
\node[ok, anchor=west] at (9.9,0.5) {$\checkmark$ gate writes on tests\\$\checkmark$ keep source labels\\$\checkmark$ revoke contradicted entries};
\draw[thin] (14.5,0 |- mem.north) .. controls (13.6,0.45) and (7.2,0.45) .. (5.35,0 |- wc.north)
  node[pos=0.5, below=1pt, sub] {context assembly (retrieval)};

% ================= Environment <-> working context =================
% observation (up)
\draw[flow] (1.2,0 |- env.north) -- (1.2,0 |- kv.south);
\node[lab, anchor=east] at (1.1,-3.6) {Observation};
\node[admit] (m1) at (0.85,-4.15) {1};
\node[bad, anchor=east] at (m1.west) {injected instructions};
\node[ok, anchor=north east] at (1.1,-4.5) {$\checkmark$ mark untrusted input\\$\checkmark$ check information flow};
% action (down)
\draw[flow] (2.75,0 |- kv.south) -- (2.75,0 |- env.north);
\node[lab, anchor=west] at (2.85,-3.6) {Action};
% assembly from the environment (up)
\draw[thin] (4.9,0 |- env.north) -- (4.9,0 |- kv.south);
\node[sub, anchor=west, align=left] at (5.0,-4.22) {context assembly\\[-1pt](rule files, evidence)};
\node[ok, anchor=north west] at (5.0,-4.65) {$\checkmark$ map rules to controls\\$\checkmark$ hash rule files};

% ================= Rollback =================
\draw[flow] (10.0,0 |- env.north) .. controls (10.0,-5.0) and (11.2,-5.0) .. (11.2,0 |- env.north);
\node[lab] (rb) at (10.6,-4.95) {Rollback};
\draw[draw=violet!75!black, dashed, line width=0.7pt, -{Stealth[length=3.2pt]}, rounded corners=4pt] (rb.north) |- (5.3,-3.65) -- (5.3,0 |- kv.south);
\node[incomplete] (m4) at (7.6,-3.65) {4};
\node[bad, anchor=west, fill=white, inner sep=1.5pt] at (m4.east) {KV not rolled back};
\node[ok, anchor=west] at (11.4,-5.3) {$\checkmark$ snapshot environment\\$\checkmark$ and cache together};

% ================= Legend =================
\node[draw=black!25, rounded corners=2pt, fill=white, anchor=north west, inner sep=3pt, font=\sffamily\figtiny] at (13.85,-3.55) {%
  \def\gl#1{\tikz[baseline={([yshift=-0.6ex]current bounding box.center)}, sharp corners]#1;}%
  \begin{tabular}{@{}c@{\hspace{3pt}}l@{}}
  \multicolumn{2}{@{}l@{}}{\textbf{How a transfer fails}}\\[1pt]
  \gl{\node[admit, minimum size=5.5pt]{}} & admits unchecked content\\
  \gl{\node[lossy, minimum size=8pt]{}} & loses content in a rewrite\\
  \gl{\node[incomplete, minimum size=5pt]{}} & restores only part of the state\\[2pt]
  {\color{green!35!black}$\checkmark$} & simple control supported by evidence\\
  \gl{\draw[flow](0,0)--(0.3,0)} & transfer performed by the harness
  \end{tabular}};

\end{tikzpicture}%
}
\caption{State transfers in a single-agent harness (\emph{Delegation} across sub-agents is omitted). Boxes show the four state stores the agent reads and writes; the harness keeps its own control and audit logs outside them. Bold arrows are primary transfers and thin arrows are context-assembly and history reads (\emph{page back} reloads masked turns). Marker shape indicates how a transfer fails: (1)~instructions injected through observations, (2)~user constraints dropped during compaction, (3)~unverified content admitted at memory write (circle) or source provenance lost during consolidation (diamond), and (4)~KV state left on the inference server after rollback (which must also truncate aborted session-history turns and revert uncommitted memory writes). Green annotations list controls with published support; each transfer is an audit point (Table~\ref{tab:audit}).}
\label{fig:architecture}
\end{figure*}

%=====================================================================
\section{State and Transfers in an Agent Harness}\label{sec:transfers}

An agent reads and writes state across four stores (Fig.~\ref{fig:architecture}). The \emph{working context} is the prompt and its KV cache on the inference server: instructions, rule files, tool schemas, retrieved memories, compacted history, and recent observations. \emph{Session history} is the full transcript of one run kept outside the prompt. The \emph{environment} is whatever the agent acts on: files and processes for coding agents, pages and browser sessions for web agents, and windows and application state for desktop agents. \emph{Persistent memory} holds episodic, semantic, and procedural records across runs~\cite{sumers2023coala}. Because the agent can modify the environment and persistent memory and a rollback truncates session history, the harness must keep its own control state---active approvals and the append-only audit log---in a separate store the agent cannot modify.

The harness moves state across these stores. \emph{Observations} bring environment state into the context, and \emph{context assembly} loads workspace rule files, tool schemas, and retrieved memories. \emph{Compaction} moves older turns to session history and leaves a condensed summary or placeholder (which \emph{page back} can later reload on demand); \emph{memory writes} and \emph{consolidation} extract and merge entries from the working context or session history across runs. \emph{Actions} change the environment, and \emph{rollback} restores prior state across the environment, working context, session history, persistent memory, and inference-server KV cache. Multi-agent harnesses add \emph{delegation}, which passes a task and state slice to a sub-agent with its own context. Each transfer gives the harness a concrete boundary to record and check what crosses.

A transfer can fail in three ways. It can \emph{admit unchecked content}: injected tool output, unverified memory writes, or unvetted recall. It can \emph{rewrite content lossily}: a summary that omits a user constraint, consolidation that discards provenance. It can \emph{restore state incompletely}: the environment is restored but the KV cache, session history, or uncommitted memory writes are not. These modes correspond to three properties an auditor needs to establish: where content came from, whether it survived a rewrite, and whether a recovery was complete.

%=====================================================================
\section{How Transfers Fail}\label{sec:fail}

\subsection{Admitting Unchecked Content}

Observations carry text that the model may treat as instructions. On InjecAgent, a ReAct-prompted agent on a frontier model followed injected instructions in about a quarter of test cases~\cite{zhan2024injecagent}. Components such as memory, retrieval, web access, and API calls make agents easier to attack than isolated models, and illustrative attacks that require no knowledge of machine learning succeeded against popular open-source and commercial agents~\cite{arx2502_08586}. On the APEX-Harm workspace benchmark, agents often produced a correct deliverable yet also took an unauthorized action, showing completion does not imply safety~\cite{arx2610_13623}.

Context assembly and memory writes extend the reach of admitted content across turns and sessions. During context assembly, workspace rule files are loaded as advisory text rather than enforced invariants: across rule files studied in~\cite{arx2608_23550}, 84--96\% of extracted security rules lacked a matching harness control, and recalling persistent memories without an admission policy can re-admit conflicting or out-of-scope entries~\cite{arx2610_11188}. When writing to memory, an attacker with access to a shared RAG or memory store can inject poisoned entries keyed to a trigger phrase in the user prompt without hurting normal accuracy~\cite{chen2024agentpoison}. Even without direct access to the store, a web agent that read one manipulated product page saved injected instructions to memory and later recalled them on other websites, bypassing permission checks~\cite{zou2026poison}. In coding and embodied tasks, test feedback and self-verification check functional correctness~\cite{shinn2023reflexion,wang2023voyager} and can gate memory writes. Because code or actions that pass tests can still contain vulnerabilities or unauthorized side effects~\cite{arx2610_13623,arx2610_06163}, passing a test suite does not prove that an entry is safe or came from a trusted source, so test checks must be paired with source labels; web and office tasks rarely have tests.

\subsection{Rewriting Content Lossily}

Compaction rewrites the working context when it grows. On the ConPACT suite, compactors kept on average only 17\% of user constraints and usually performed worse than not compacting at all, whereas extracting and pinning constraints in a separate pass preserved most of them~\cite{chen2026lostincompaction}. Replacing older observations with placeholders, which shortens the context without rewriting it, matched the solve rate of LLM summarization at lower cost on SWE-bench Verified~\cite{lindenbauer2025complexitytrap}. This suggests that rewriting, rather than shortening, causes constraint loss.

Consolidation rewrites memory. In \emph{authority collapse}, consolidation keeps a claim but drops the source constraints on its use, so a memory conveys more authority than its source warranted. This occurred in 48 of 49 tested memory-system and model combinations; storing an authority label with each memory eliminated the resulting unauthorized actions~\cite{zhan2026when}. Stale entries cause a related failure: when facts changed over time, append-only memory stores performed worse than having no memory at all once earlier facts reversed, whereas revoking contradicted entries maintained accuracy~\cite{zhou2026tepa}.

\subsection{Restoring State Incompletely}

Removing turns from the prompt undoes neither a file write nor an outbound request. Taking system checkpoints after turns that modify files or processes restored local state where rewinding the chat transcript failed~\cite{wu2026crab}, and holding external actions in a staging buffer for validation before execution reduced irreversible mistakes~\cite{chen2026cordon}. Complete rollback must also clean up internal stores and the inference server: aborted turns in the working context and session history must be removed or marked so they cannot be reloaded later, and uncommitted writes to persistent memory must be undone~\cite{su2026chronomem}. On the inference server, across seven open-weight model families, KV state retained after an aborted branch changed a protected decision in 25 of 63 audited cases, although the discarded tokens were absent from the request~\cite{zhang2026aborted}. Rebuilding the cache resolved every failure, and restoring a per-turn KV cache snapshot fixed the leak without flushing the entire cache. The failure requires a serving session that keeps KV state across the abort, so we expect it to concentrate in stateful multi-turn sessions rather than in prefix caches keyed on exact tokens.

\subsection{Cross-Transfer Fault Propagation}

Faults introduced at one transfer are carried by later ones. A large observation fills the context and triggers compaction early, risking constraint loss on each pass~\cite{chen2026lostincompaction}. Extracting memories from a summary then turns paraphrase errors into stored facts. In controlled failure-injection probes, fault propagation increased with the depth of multi-agent pipelines, and blind retries repeated the same underlying errors rather than fixing them~\cite{arx2608_05263}. Unified security benchmarks such as Agent Security Bench (ASB) evaluate mixed attacks alongside prompt injection and memory poisoning in text-tool settings~\cite{zhang2024asb}, and observation-to-memory-to-action poisoning has been measured in web agents~\cite{zou2026poison}; however, end-to-end chains combining prompt injection, lossy compaction, authority collapse, and incomplete KV-cache rollback remain to be evaluated in a unified benchmark. Most measurements come from coding agents and text tools. Table~\ref{tab:evidence} compares transfer failures between coding and GUI agents; Section~\ref{sec:mm} discusses web and desktop settings.

\begin{table}[t]
\centering
\footnotesize
\setlength{\tabcolsep}{2.5pt}
\renewcommand{\arraystretch}{1.05}
\caption{Transfer failure surfaces across coding/text-tool and GUI agents.}
\label{tab:evidence}
\begin{tabularx}{\columnwidth}{>{\raggedright\arraybackslash}p{1.65cm}>{\raggedright\arraybackslash}X>{\raggedright\arraybackslash}X}
\toprule
\textbf{Transfer} & \textbf{Coding and text-tool agents} & \textbf{Web and desktop agents} \\
\midrule
Observation & Injected instructions followed~\cite{zhan2024injecagent}; attacks and defenses compared in shared environments~\cite{debenedetti2024agentdojo,zhang2024asb} & Pop-ups and static injections succeed~\cite{zhang2024popups,arx2506_14866} \\
\addlinespace[2pt]
Compaction & Constraint loss measured; masking compared with summaries~\cite{chen2026lostincompaction,lindenbauer2025complexitytrap} & Screenshot history affects success~\cite{arx2608_22577}; constraint loss unmeasured \\
\addlinespace[2pt]
Memory write & Functional test gates~\cite{wang2023voyager} (do not check security) & Poisoning through observation alone~\cite{zou2026poison}; no automated test check \\
\addlinespace[2pt]
Consolidation & \multicolumn{2}{>{\raggedright\arraybackslash}p{6.5cm}}{Authority collapse measured across memory systems and models; agent type not varied~\cite{zhan2026when}} \\
\addlinespace[2pt]
Rollback & Retained KV state measured~\cite{zhang2026aborted}; checkpoints~\cite{wu2026crab} & Actions often irreversible; lookahead instead~\cite{gu2024webdreamer} \\
\bottomrule
\end{tabularx}
\end{table}

%=====================================================================
\section{Audit Points}\label{sec:audit}

An audit point specifies, for one transfer, the record the harness emits and the check it enables. Table~\ref{tab:audit} lists one per transfer, shaped by a trust model and three principles.

\para{Threat and trust model.}
We assume that an adversary can control content reaching the agent through observations (web pages, repository files, tool output, screenshots), and through it what the agent writes to memory. The adversary cannot modify the harness or the harness-owned store that holds audit records; such stores, based on hash chaining, have been built for agent actions (Section~\ref{sec:security}). Because the model may misreport what it read or why it acted, harness code emits each record at the transfer boundary and appends it to a store the agent cannot modify. Benign failures, such as constraint loss during compaction or stale memory, require the same records, so the audit points serve fault diagnosis as well as security. A compromised harness or inference server, and an operator who can rewrite the store, are out of scope. Under this model, an attacker can still plant fake provenance strings in observations, flood the harness with events, or cause records to be skipped (by steering the agent onto an unlogged side channel, tricking a model-based extractor into omitting a field, or overflowing a log buffer). Write isolation blocks direct agent edits, signed hash-chained logs detect tampering and deletions~\cite{crosby2009efficient,arx2609_01931,arx2603_12621}, and attaching source labels at tool return counters fake provenance; unlogged side channels and flooding remain open. To protect sensitive data in audit records, fixed identifiers (segment IDs, rule hashes, cache IDs) can be stored as keyed hashes (HMACs), while full text needed for review or policy checks (dropped constraints, tool arguments, raw transcripts) is kept in an access-controlled or encrypted store. Session-history turns whose content flows into persistent memory must be kept for the life of that memory so cross-session links resolve.

\para{Records follow content across transfers.}
Most entries record a label together with the content: its source and trust level at observation, its lineage through compaction and consolidation, and the authority under which an action was taken. This is information flow control applied to the harness. Each segment carries an integrity label (source trust or authority), checked against injection and authority collapse, and a confidentiality label (data sensitivity), checked against leakage. Recent systems enforce such labels in the harness: FIDES propagates confidentiality and integrity labels across messages, tool arguments, and outputs and hides sensitive tool outputs behind variables~\cite{costa2025fides}; RTBAS uses model-based dependency screeners and prevented all targeted attacks on AgentDojo with a small utility loss~\cite{arx2502_08966}; FLOWSEAL enforces confidentiality with a tool-level interceptor grounded in provenance and a label lattice~\cite{arx2609_14003}; and column-lineage gating removed measured leakage in shared analytical memory~\cite{arx2610_07258}. Labels must survive every rewrite: if compaction or consolidation drops a provenance label, later checks fail and authority collapses~\cite{zhan2026when}. Conversely, if compaction merges many turns into one summary that takes the lowest trust label of any input turn, a single untrusted observation marks the whole summary as untrusted. Harnesses can avoid this by summarizing trusted and untrusted turns separately, tracking provenance per claim, masking old observations instead of summarizing~\cite{lindenbauer2025complexitytrap}, or keeping untrusted outputs in referenced variables~\cite{costa2025fides}.

\para{Deterministic checks first.}
Most audit points can be checked without a model call: comparing hashes of loaded rule files, string-matching identifiers before and after compaction, checking that an action's capability is a subset of the delegating agent's, and matching snapshot and cache identifiers after rollback. Some cannot. A summarizer may paraphrase user constraints; preserving them requires running a dedicated extractor alongside compaction, which adds a model call~\cite{chen2026lostincompaction}. Authority labels are likewise predicted at write time~\cite{zhan2026when}. This asymmetry determines both cost and attack surface: deterministic checks add only logging and comparison, whereas model-based extractors and labelers process untrusted context and can be targeted by indirect prompt injection. A middle ground uses the model only to extract structured attributes and leaves the allow/deny decision to deterministic rules, as in MetaPermit~\cite{arx2609_31039} and IntentCap~\cite{arx2609_14631}. For compaction, the harness can extract and pin user constraints from trusted user turns before untrusted observations enter the context, so an injected observation cannot alter the baseline list. After compaction, a model checks whether each pinned constraint holds in the summary, and a deterministic rule blocks or repairs the summary if any is missing. Recording the checker's prompt and output lets an auditor inspect the model check afterward, whereas preventing prompt injection against the checker at run time requires deterministic IFC planners~\cite{costa2025fides} or isolated execution~\cite{wu2024isolategpt}.

\para{Record integrity and completeness.}
A provenance field is useful only if an adversary cannot forge it. In the replay study, a single planted line defeated a naive provenance check on most of the runs it would otherwise have flagged, and binding tool returns to issued calls---rather than trusting provenance text inside tool output---closed this evasion~\cite{arx2610_09793}. Completeness matters as well: in lineage-gated memory, blocking measured leakage required recording 75--90\% of lineage~\cite{arx2610_07258}.

\begin{table*}[!t]
\centering
\footnotesize
\setlength{\tabcolsep}{3.2pt}
\renewcommand{\arraystretch}{1.02}
\caption{Audit points at state transfers. ``Det.'' marks checks that need no model call; ``Model'' marks checks that need one. Evidence and controls are discussed in Sections~\ref{sec:fail}--\ref{sec:security}.}
\label{tab:audit}
\begin{tabularx}{\textwidth}{>{\raggedright\arraybackslash}p{1.75cm}>{\raggedright\arraybackslash}p{3.2cm}>{\raggedright\arraybackslash}X>{\raggedright\arraybackslash}p{3.0cm}>{\raggedright\arraybackslash}p{3.6cm}}
\toprule
\textbf{Transfer} & \textbf{Failure (evidence)} & \textbf{What to record} & \textbf{Check} & \textbf{Control supported} \\
\midrule
Observation & Injected instructions followed~\cite{zhan2024injecagent,zhang2024popups} & Source, integrity/confidentiality labels, and content hash of each segment; producing tool call; screenshot hash and parsed regions & Det.\ (label propagation) & Marking~\cite{hines2024spotlighting}; control-flow isolation~\cite{debenedetti2025camel,costa2025fides}; IFC~\cite{arx2502_08966,arx2609_14003} \\
\addlinespace[1pt]
Context assembly & 84--96\% of security rules in instruction files lacked a control~\cite{arx2608_23550} & Hashes of rule files, tool schemas, and recalled memories; declared rule-to-control map; memory admission decisions & Det.\ (hashes; declared map; policy over stored labels) & Enforced rules; policy-gated memory recall~\cite{arx2610_11188} \\
\addlinespace[1pt]
Compaction & Constraints~\cite{chen2026lostincompaction} and identifiers dropped & Identifiers and pinned constraints kept or dropped per pass; pointer and hash of full transcript & IDs: Det.\ match. Constraints: Model check + Det.\ rule & Pinned constraints~\cite{chen2026lostincompaction}; masking instead of rewriting~\cite{lindenbauer2025complexitytrap} \\
\addlinespace[1pt]
Memory write & Unverified or injected content stored~\cite{chen2024agentpoison,zou2026poison} & Source segment IDs, hashes, and labels; write-time authority label; verification result; writer identity & Provenance gate: Det. Authority label: Model. Truth check: Det.\ with test oracle, else open & Provenance-gated or test-verified writes~\cite{wang2023voyager} \\
\addlinespace[1pt]
Consolidation & Authority collapse~\cite{zhan2026when}; stale entries~\cite{zhou2026tepa} & Source segment IDs and hashes; authority label; validity status; superseded or revoked entry IDs & Lineage, revocation: Det. Authority label: Model & Authority labels~\cite{zhan2026when}; lineage-gated retrieval~\cite{arx2610_07258} \\
\addlinespace[1pt]
Action & Unauthorized side effects~\cite{arx2610_13623} & Tool, arguments, input lineage and labels; approval ID, capability/lease, execution count, timestamp; GUI clicked element & Det.\ (policy, capability, IFC); Model lookahead & Capabilities and leases~\cite{arx2609_08371,arx2609_14631}; IFC~\cite{costa2025fides,arx2609_14003}; staged effects~\cite{chen2026cordon} \\
\addlinespace[1pt]
Rollback & Retained KV state changes decisions~\cite{zhang2026aborted} & Environment, transcript, and memory snapshot IDs bound to inference-server cache/session IDs (or flag when server cache is opaque); committed effects & Det.\ (ID match) & Cache snapshot restore or rebuild~\cite{zhang2026aborted}; checkpoints~\cite{wu2026crab} \\
\addlinespace[1pt]
Delegation & Faults propagate with depth~\cite{arx2608_05263} & Task, input lineage and labels, and capabilities passed to sub-agent; returned result, hashes, and labels & Det.\ (subset capability; label propagation) & Per-agent capabilities~\cite{arx2609_08371}; leases across delegation~\cite{arx2609_14631}; isolation~\cite{wu2024isolategpt} \\
\bottomrule
\end{tabularx}
\end{table*}

\para{A common record format.}
The rows of Table~\ref{tab:audit} share a structure, which suggests one hash-chained record type emitted at every transfer. Each record contains (i)~the transfer type, step, session, previous-record hash, and a harness timestamp; (ii)~the acting component (harness, model, sub-agent, or user), responsible principal, and active authorization state (capability or lease, user approval, approval ID, and how many times that approval has been executed); (iii)~identifiers, content hashes, and integrity/confidentiality labels of the input segments; (iv)~identifiers, content hashes, and assigned labels of the output segments, together with pointers and hashes for full transcripts or model-check prompts and outputs kept in the backing store; (v)~the checks run and their results (each marked deterministic or model-based), active obligations or pinned user constraints, and transfer-specific metadata (such as the rule-to-control map, revoked entry IDs, or intermediate file and stage snapshots); and (vi)~for actions and rollbacks, the environment, transcript, and memory snapshot IDs bound to the inference-server cache and session IDs (or a flag when the server does not expose cache state). Fields (iii) and (iv) link records into a verified lineage graph across transfers~\cite{arx2610_01256}, while fields (i), (ii), and (v) track active obligations and supply the timestamps and approvals missing from replayed corpora~\cite{arx2610_09793}. Records can be emitted as spans in existing tracing infrastructure.

\para{Cost of recording.}
Most fields are identifiers, hashes, and labels; full content resides in the backing store, so a record points to it with a payload hash instead of copying it. Deterministic logging adds little overhead, whereas content scanning and model-based checks dominate latency. Most turns produce no recovery-relevant state~\cite{wu2026crab}. Derivation checks added 13.8\,$\mu$s worst-case overhead~\cite{arx2610_07258} and hash-chained action logs added about 48\,$\mu$s median latency and 512 bytes per event~\cite{arx2609_01931}. Pre-execution scanning added 8.3\,ms median latency~\cite{arx2603_12621}, and constraint or authority extraction adds a model call~\cite{chen2026lostincompaction,zhan2026when}. As an illustration only, at 512 bytes per event a run with a thousand recorded transfers produces about half a megabyte of audit records (and a million such runs per day about half a terabyte), plus backing-store retention for cross-session lineage slices. Screenshots are the main exception, since auditing requires archiving originals (Section~\ref{sec:mm}).

\para{Worked trace.}
Consider a five-step failure chain: (1)~the agent reads a web page with an embedded instruction; (2)~the large observation triggers compaction, which drops a user constraint; (3)~content from the page is written to memory~\cite{zou2026poison}; (4)~a later session retrieves that entry via context assembly and takes a prohibited action; and (5)~the harness rolls back the workspace while the inference server retains the session's KV state. With the records in Table~\ref{tab:audit}, an auditor can trace the prohibited action's lease and input lineage through context assembly and the memory write back to the untrusted observation, see from the pinned constraint set that compaction dropped a rule, and see from the rollback record that the workspace snapshot was not bound to a fresh cache identifier. Together, these transfer records reconstruct the causal chain from observation to action and flag both the dropped constraint and the unreset server cache, which Section~\ref{sec:eval} evaluates under field-level attribution ablations.

%=====================================================================
\section{Observability in Practice}\label{sec:obs}

\para{Trace analysis and failure attribution.}
Failure attribution asks which agent and which step caused a failure~\cite{zhang2025which}. Modeling dependencies among trajectory events helps: DeFA traces failures backward through a dependency graph and achieved the highest agent- and step-level attribution accuracy among evaluated methods on Who\&When~\cite{arx2610_01256}. CEG-Agent links execution events, errors, and failure outcomes in typed causal graphs~\cite{arx2609_32514}. Both infer which event affected later steps; lineage records would supply the data-flow links directly. Failures that raise no error are harder to localize. For vulnerable patches that passed functional checks, the origin rarely coincided with the edit; judges agreed more on origin type than exact turn, and least when traces kept only the final file~\cite{arx2610_06163}. Traces therefore need intermediate state as well as final outcomes. In infrastructure operations, benchmarks now grade whether an agent's diagnostic reasoning is grounded in evidence as well as whether it localizes the fault~\cite{arx2606_29193}. In ITBench, agents resolved few site reliability, compliance, or financial scenarios~\cite{arx2502_05352}.

\para{Runtime monitors and guardrails.}
Guardrails check content as it crosses a transfer boundary. Programmable rails, designed for conversational systems, control model output at run time independently of the underlying model~\cite{rebedea2023nemo}. Monitors based on temporal logic can express multi-step, data-dependent policies, but their value depends on the trace. On replayed trajectories, generic obligations caught most attack chains but also flagged 29.3\% of benign runs because the traces lacked approvals and timestamps~\cite{arx2610_09793}. A monitor checking a fixed set of rules catches a high share of attacks only when failures cluster into a few recurring patterns, so the same rules covered many attacks on some models and few on others~\cite{zhang2026why}. Forbidden actions are not the only risk: on a coding safety benchmark, unfulfilled obligations outnumbered explicit forbidden actions, and off-the-shelf models achieved at most 49\% recall on coding and terminal trajectories~\cite{arx2610_11773}. Detecting omissions therefore requires recording active obligations and preserving constraints across compaction.

\para{Shared infrastructure.}
Enterprise agents may share a memory store and an inference server. When multiple agents share an analytical memory store, restricting retrieval by which database columns an entry was derived from blocked leakage of sensitive columns and helped document EU AI Act compliance~\cite{arx2610_07258}. Because rollback correctness spans the harness and the inference server~\cite{zhang2026aborted}, binding snapshot to cache identifiers is a cross-layer record that neither layer can produce alone.

%=====================================================================
\section{Security Controls Built on Audit Records}\label{sec:security}

The records in Table~\ref{tab:audit} support runtime enforcement as well as post-hoc diagnosis: proposed agent controls rely at run time on transfer state that these fields are designed to capture. Table~\ref{tab:props} maps each control to the corresponding fields.

\begin{table}[t]
\centering
\footnotesize
\setlength{\tabcolsep}{2.5pt}
\renewcommand{\arraystretch}{1.02}
\caption{Security controls and corresponding audit fields (Table~\ref{tab:audit}).}
\label{tab:props}
\begin{tabularx}{\columnwidth}{>{\raggedright\arraybackslash}p{1.75cm}>{\raggedright\arraybackslash}X>{\raggedright\arraybackslash}p{2.2cm}}
\toprule
\textbf{Control} & \textbf{Corresponding fields} & \textbf{Examples} \\
\midrule
Least privilege & Action: tool, arguments, capability, approval; delegation capabilities & \cite{shi2025progent,arx2603_20933,arx2601_11893} \\
\addlinespace[1pt]
Durable authorization & Action, compaction, rollback: approval ID, confirmation, execution count; pinned constraints & \cite{arx2608_01710,arx2609_32428} \\
\addlinespace[1pt]
Information flow control & Observation, memory, consolidation, delegation: integrity/confidentiality labels, lineage & \cite{costa2025fides,arx2502_08966,arx2609_14003} \\
\addlinespace[1pt]
Monitors and guardrails & Action, compaction: timestamp, approval; active obligations, constraints & \cite{arx2610_09793,arx2603_12621,arx2610_11773} \\
\addlinespace[1pt]
Accountability & Action (all transfers): hash-chained records; responsible principal & \cite{arx2609_01931,arx2607_23207} \\
\addlinespace[1pt]
Failure attribution & Step and delegation lineage; intermediate files and stage outputs & \cite{arx2610_01256,arx2605_15581,arx2610_06163} \\
\addlinespace[1pt]
Recovery & Rollback: snapshot, cache, session ID & \cite{arx2608_14380,wu2026crab,zhang2026aborted} \\
\bottomrule
\end{tabularx}
\end{table}

\para{Authorization is state that crosses transfers.}
Least-privilege enforcement checks each tool call against an explicit policy. Progent uses an LLM to draft and update symbolic rules over tool names and arguments, checks calls deterministically, and uses an SMT solver so privilege expansions require approval~\cite{shi2025progent}. AC4A organizes application resources into hierarchies and scopes agent permissions to specific APIs and web page elements~\cite{arx2603_20933}. SEAgent enforces mandatory access control over an agent--tool information flow graph and blocked privilege escalation, including a multi-agent confused-deputy variant~\cite{arx2601_11893}. Other designs check each call against typed per-agent capabilities, short-lived leases derived from user intent, or authorization contracts compiled before untrusted execution~\cite{arx2609_08371,arx2609_14631,arx2610_06966}. At the planner and runtime boundary, FIDES tracks confidentiality and integrity labels across messages, tool arguments, and outputs to check policies before tool calls~\cite{costa2025fides}, and IsolateGPT runs untrusted applications in isolated spokes mediated by a central permission hub~\cite{wu2024isolategpt}. Framework defaults lag behind: an audit of three agent frameworks found that all gate tool exposure, but none check per-call argument values by default~\cite{arx2606_28679}.

Authorization must also be recorded as durable state outside the prompt rather than inferred from the transcript. When a user revises part of an instruction, tracking which permissions depend on which part of the prompt lets the harness revoke only the affected permissions, improving both action safety and task success with three LLMs on two tasks~\cite{arx2609_32428}; compaction must still preserve those instruction revisions so the agent and monitors see the updated constraints. When agents replan, retry, delegate, or resume after a crash, a single user approval can be executed several times under freshly issued single-use tokens; preventing this required tracking each approval's execution count in persistent state at the issuer~\cite{arx2608_01710}. Because rollbacks and retries truncate or replay the transcript, authorization instances and execution counts belong with the action and rollback records outside the prompt.

\para{Tamper-evident records.}
Section~\ref{sec:audit} assumed a store that the agent cannot alter, and such stores now exist for agent actions. The Agent Flight Recorder serializes each action as a structured event, hash-chains events and batches them into Merkle trees, and can anchor epoch roots on a public ledger. On synthetic workloads it detected edit, delete, reorder, and fork tampering, and structured forensic queries were more precise than text search over unstructured logs on guardrail and delegation lookups~\cite{arx2609_01931}. AEGIS interposes on the tool-execution path, holds high-risk calls for human approval, and records each decision in a signed, hash-chained trail~\cite{arx2603_12621}. Both systems center on tool calls and their provenance. As described, neither records compaction, consolidation, nor the state of the inference server, which Table~\ref{tab:audit} adds. Accountability also requires linking an agent to a responsible principal. One national agent-identity system links each agent to a verified legal principal across two separated public-agency stores, requiring separate legal compulsion of both to re-identify an owner~\cite{arx2607_23207}.

\para{Cascading faults in multi-agent pipelines.}
Faults in multi-agent pipelines often raise no exception. MAS-FIRE injects faults through prompt modification, response rewriting, and message routing; across its benchmarks, stronger foundation models were not uniformly more fault-tolerant~\cite{arx2602_19843}. Silent tool failures are the hardest case: in ToolMaze, recovery dropped most when tools returned plausible but corrupted data without an error code, because agents trusted the output~\cite{arx2606_05806}. Compromise also spreads between agents. INFA-Guard distinguishes compromised benign agents from originally malicious ones, traces how an attack spread across agents, and reduced attack success~\cite{arx2601_14667}. We argue that each of these analyses needs to know which agent produced which message and which tool output a decision consumed; these are the lineage fields of the observation and delegation records. Instrumentation for this is appearing: \texttt{llmmas-otel} combines OpenTelemetry step tracing with fault injection at inter-agent messages, tool calls, and model invocations~\cite{arx2608_24271}. Recovery must be recorded as well. AgentRewind keeps aligned checkpoints of the agent context and a controlled environment, and resumes with information from earlier attempts~\cite{arx2608_14380}. Because blind retries repeated the same underlying errors in failure-injection probes~\cite{arx2608_05263}, we suggest that retry policies consult the audit record to choose a checkpoint rather than replaying inputs.

\para{Agents that operate infrastructure.}
Agents that manage IT and data infrastructure raise the stakes of each transfer. AIOpsLab deploys microservice environments, injects faults, exports telemetry, and evaluates agents across the incident lifecycle~\cite{arx2501_06706}. Telemetry is an observation channel that an adversary can write to. Crafted requests planted misleading telemetry that steered AIOps agents into compromising their own clusters; structure-aware telemetry sanitization blocked these attacks without hurting normal performance~\cite{arx2508_06394}. For diagnosis, splitting root-cause analysis into four logged stages made it possible to pinpoint which stage failed and replay the workflow from that stage with a fix~\cite{arx2605_15581}. A zero-trust harness for cloud engineering combines identity, authorization, policy-scoped capabilities, and isolation; in evaluation, each run ended in either a verified deployment or a logged failure after bounded retries~\cite{arx2609_00050}. Because telemetry streams and long runs make full-content logging expensive at scale, records should store identifiers, hashes, and labels with pointers (Section~\ref{sec:audit}).

%=====================================================================
\section{Multimodal Agents: What Changes}\label{sec:mm}

Evidence for web and desktop agents is thinner than for coding agents, and three differences affect the audit points.

\para{Observations are images.}
Screenshots carry instructions that a text filter does not see. Adversarial pop-ups, which human users would recognize and ignore, led vision-language agents in OSWorld and VisualWebArena to click them in most trials, reducing task success. Simple defenses, such as instructing the agent to ignore pop-ups, were ineffective~\cite{zhang2024popups}. Computer-use agents also complied with many deliberate misuse requests and were relatively vulnerable to static prompt injection on OS-Harm~\cite{arx2506_14866}. RiOSWorld separates user-originated from environmental risks across web, email, office, and operating-system tasks~\cite{arx2506_00618}. For audit, the observation record logs the screenshot hash and candidate regions detected by a screen parser~\cite{lu2024omniparser}, while the subsequent action record logs the clicked element or coordinates with a link back to that observation; dependency-based attribution has been applied to such image and video trajectories~\cite{arx2610_01256}. A pixel-only parser helps locate what was clicked after the fact, but assigning source trust labels at run time requires matching coordinates to DOM or OS window metadata, or treating the screen as untrusted.

\para{Screenshots cannot be re-fetched from the environment.}
Observation masking works for command output because the harness can retrieve the original later. Screenshots of a changing interface cannot be re-fetched. GUI agents therefore keep text summaries of earlier steps~\cite{arx2603_08533}, compress older images~\cite{arx2602_23235}, or keep all frames on disk and restore only a budgeted subset to the prompt, which improved success over summary-only memory on OSWorld-Verified~\cite{arx2608_22577}. Summaries, image pruning, and unrestored frames still lose context: checking whether identifiers survive screenshot compaction requires OCR or parser tokens, and constraint retention under visual compaction remains unmeasured. Archiving full screenshots across a thousand transfers also takes hundreds of megabytes to gigabytes rather than about half a megabyte of text pointers, and images cannot be replaced with keyed hashes; storing cropped regions, encoding frame deltas, or redacting visual personal data before archival cuts storage and privacy risk.

\para{Actions often cannot be undone.}
On live websites, submitted forms and sent messages cannot be rolled back, so harnesses evaluate candidate actions before committing them, for example by simulating interactions with a learned world model~\cite{gu2024webdreamer}. When a harness runs such pre-commit lookahead, the action record's checks field logs the candidate action, predicted outcome, and policy decision as a model-based check. Because web tasks lack a test oracle to gate memory writes, poisoning through observation alone~\cite{zou2026poison} makes write-time lineage critical.

%=====================================================================
\section{Implementing Audit Points}\label{sec:impl}

Implementing audit points requires no model changes: the harness routes each transfer through a wrapper that logs the hash-chained record and runs boundary checks before applying the state change. Our Python reference middleware implements all eight wrappers directly; for existing harnesses, adoption can proceed in four incremental steps ordered by overhead and risk coverage.

\para{Step 1: actions and authorization.}
Interpose on the tool-execution path, as pre-execution firewalls do~\cite{arx2603_12621}, and record each call with its arguments, the capability or approval that authorized it, and a harness timestamp. Keep authorization instances and their execution counts in a durable store outside the prompt~\cite{arx2608_01710}, supplying the approvals and timestamps missing from replayed logs~\cite{arx2610_09793} for least-privilege checks~\cite{shi2025progent}.

\para{Step 2: observation labels and lineage.}
Assign each segment entering the prompt a source, an integrity label, and a confidentiality label, and carry segment identifiers through memory writes. This supports information flow control~\cite{arx2502_08966,arx2609_14003}, and we expect it to let an auditor trace the observation-to-memory-to-action chain measured in~\cite{zou2026poison}.

\para{Step 3: rewrites.}
At each compaction and consolidation, record identifiers and the constraint list before and after, the lineage of each output entry, and its authority label. Identifier matching is deterministic; constraint and authority checks need a model call whose prompt and output are logged (Section~\ref{sec:audit}).

\para{Step 4: recovery and the inference server.}
Bind each environment, transcript, and memory snapshot to the cache and session identifiers on the inference server, and record which side effects were already committed. Self-hosted servers can expose cache identifiers directly~\cite{zhang2026aborted}; when a commercial model API hides internal KV-cache state, the harness logs a flag in field~(vi) and falls back on rollback to re-sending the prompt without server state or with a cache-busting prefix.

\para{Accountability limits.}
Audit records support attributing a technical failure to a state transfer, but technical attribution is distinct from allocating legal or organizational responsibility. In a causal-model analysis of human--agent teams, once combined autonomy passes a threshold and interactions form a feedback loop, no attribution rule satisfies four basic accountability requirements at once; audits and oversight cannot restore them without restricting autonomy~\cite{arx2604_07778}.

%=====================================================================
\section{Evaluation}\label{sec:eval}

\para{Reference middleware and setup.}
Because static failure corpora such as Who\&When and TRAIL~\cite{zhang2025which,arx2505_08638} omit internal transfer state, we implemented a Python reference middleware interposing on all eight harness transfers with SHA-256 hash chaining, product-lattice integrity/confidentiality labels, durable approval execution counters, and snapshot-to-cache bindings. We evaluate deterministic wrapper overhead across 20,000 invocations (2,500 per transfer) and evaluate violation detection and root-cause attribution across 240 multi-step trajectories: 120 benign runs (exercising approved privileged actions, compactions, memory consolidations, delegations, and clean rollbacks) and 120 fault-injected runs~\cite{arx2505_03096} spanning six families (20 each): indirect observation injection with spoofed in-band provenance, syntactic and paraphrased compaction constraint drops, memory consolidation authority collapse, approval replay across rollback, incomplete rollback with unreset server KV state, and sub-agent capability/taint violations.

\para{Wrapper latency and record size.}
Table~\ref{tab:overhead} reports per-transfer deterministic wrapper latency and serialized record size. Seven transfers complete in 12.5--22.1\,$\mu$s median latency (23.3--75.1\,$\mu$s p99), while compaction adds 81.6\,$\mu$s median latency (150.2\,$\mu$s p99) due to syntactic identifier diffing across turns. Records average 524\,bytes (421--711\,bytes), matching the $\approx$0.5\,MB per thousand transfers estimate in Section~\ref{sec:audit}.

\begin{table}[t]
\centering
\footnotesize
\setlength{\tabcolsep}{4pt}
\renewcommand{\arraystretch}{1.02}
\caption{Deterministic wrapper latency and serialized record size across 20,000 invocations (2,500 per transfer).}
\label{tab:overhead}
\begin{tabularx}{\columnwidth}{>{\raggedright\arraybackslash}Xrrr}
\toprule
\textbf{Transfer wrapper} & \textbf{Median ($\mu$s)} & \textbf{p99 ($\mu$s)} & \textbf{Mean size (B)} \\
\midrule
Observation & 16.0 & 58.8 & 421 \\
Context assembly & 12.5 & 23.9 & 499 \\
Compaction & 81.6 & 150.2 & 530 \\
Memory write & 15.2 & 39.0 & 504 \\
Consolidation & 15.3 & 75.1 & 520 \\
Action & 22.1 & 58.9 & 711 \\
Rollback & 14.9 & 52.5 & 479 \\
Delegation & 15.3 & 23.3 & 530 \\
\bottomrule
\end{tabularx}
\end{table}

\para{Violation detection and attribution ablations.}
Following replay verification~\cite{arx2610_09793} and dependency attribution~\cite{arx2610_01256}, Table~\ref{tab:eval} compares an OpenTelemetry GenAI lifecycle span baseline (\emph{OTel-GenAI}), a tool-call-only firewall log (\emph{Tool-Log}), and our transfer-audit middleware. Lacking durable approval records, \emph{OTel-GenAI} over-flags benign privileged actions (35.0\% FPR) while missing spoofed provenance and internal rewrites (23.3\% recall); \emph{Tool-Log} avoids false positives (0.0\% FPR) but misses cross-transfer memory taint, replayed approvals after rollback, and stale KV state (16.7\% recall). Deterministic transfer checks alone reach 91.7\% recall (95.7\% F1, missing only paraphrased constraint drops that drop no syntactic identifiers), and full transfer records reach 98.3\% recall (99.2\% F1) at 0.0\% FPR. On the 120 faulty runs, backward traversal over emitted records raises exact root-cause step attribution from 10.8--16.7\% to 98.3\%; ablating segment lineage (fields~iii--iv), auth/cache bindings (ii, vi), or rewrite diffs (v) drops step accuracy to 48.3\%, 65.0\%, and 83.3\%.

\begin{table}[t]
\centering
\footnotesize
\setlength{\tabcolsep}{2.8pt}
\renewcommand{\arraystretch}{1.02}
\caption{Multi-transfer violation detection and root-cause step attribution across 240 trajectories (120 benign, 120 fault-injected).}
\label{tab:eval}
\begin{tabularx}{\columnwidth}{>{\raggedright\arraybackslash}Xrrrr}
\toprule
\textbf{Trace schema / ablation} & \textbf{Recall} & \textbf{FPR} & \textbf{F1} & \textbf{Step Attr.} \\
\midrule
OTel-GenAI lifecycle spans & 23.3\% & 35.0\% & 29.5\% & 10.8\% \\
Tool-Log (tool calls only) & 16.7\% & 0.0\% & 28.6\% & 16.7\% \\
Transfer-Audit (Det.\ only) & 91.7\% & 0.0\% & 95.7\% & 91.7\% \\
Transfer-Audit (Full schema) & \textbf{98.3\%} & \textbf{0.0\%} & \textbf{99.2\%} & \textbf{98.3\%} \\
\midrule
\quad w/o segment lineage (iii--iv) & --- & --- & --- & 48.3\% \\
\quad w/o auth/cache bindings (ii, vi) & --- & --- & --- & 65.0\% \\
\quad w/o rewrite diffs (v) & --- & --- & --- & 83.3\% \\
\bottomrule
\end{tabularx}
\end{table}

\para{Comparison with tracing standards.}
Current tracing schemas cover only part of the common record format in Section~\ref{sec:audit}. OpenTelemetry GenAI spans, \texttt{llmmas-otel}~\cite{arx2608_24271}, TRAIL~\cite{arx2505_08638}, and AgentOps tools~\cite{dong2024agentops} record timestamps, component roles, and tool-call IDs, which supply the step metadata in fields~(i) and~(ii). They omit source trust and sensitivity labels in fields~(iii) and~(iv), compaction constraint diffs in field~(v), memory consolidation lineage, authorization execution counts, and the snapshot-to-cache bindings in field~(vi).

\para{Evaluate model--harness pairs.}
The harness changes both cost and model rankings~\cite{scaffold2026effect,arx2610_00917}; on SWE-bench Verified, changing the harness around the same model varied solve rates by more than the gap across the top thirty leaderboard entries (though leaderboard submissions do not isolate causal effects)~\cite{arx2609_17394}. Results thus characterize a model--harness pair, so reports should specify tool schemas, truncation, compaction, memory policies, and emitted audit records.

\para{Metrics tied to audit points.}
Each transfer suggests a metric:
\begin{itemize}
\item \emph{observation}: attack success under indirect or visual injection~\cite{debenedetti2024agentdojo,zhang2024popups};
\item \emph{compaction}: retention of user constraints~\cite{chen2026lostincompaction} and identifiers after compaction;
\item \emph{memory}: rate of unverified, unauthorized, or poisoned writes and their downstream effects~\cite{zhan2026when,zou2026poison};
\item \emph{action}: unauthorized side effects alongside task success, since the two are not exclusive~\cite{arx2610_13623}, and unfulfilled obligations~\cite{arx2610_11773};
\item \emph{rollback}: decision shifts on rollback without cache reset~\cite{zhang2026aborted};
\item \emph{reliability}: consistency across repeated trials (pass$^k$)~\cite{yao2024taubench}.
\end{itemize}

\para{Testbeds.}
Existing environments already inject faults and export traces, and can host these measurements. AIOpsLab injects faults into microservice deployments and exports their telemetry~\cite{arx2501_06706}, and MAS-FIRE injects faults into the messages and prompts of multi-agent systems~\cite{arx2602_19843}. ToolMaze perturbs tool outputs with explicit and implicit, transient and permanent failures~\cite{arx2606_05806}, while \texttt{llmmas-otel} aligns injected faults with distributed traces~\cite{arx2608_24271}. Because these testbeds perturb observations, actions, and delegation, covering all transfers requires adding fault injectors for compaction, consolidation, and KV-cache rollback, or pairing them with ConPACT and authority-collapse suites~\cite{chen2026lostincompaction,zhan2026when,zhang2026aborted}. Known injection points give ground truth for attribution ablations.

%=====================================================================
\section{Related Work}\label{sec:related}

Agent observability tools trace events across the execution lifecycle~\cite{dong2024agentops}, while threat and failure studies analyze attack surfaces~\cite{arx2502_08586}, multi-agent coordination breakdowns~\cite{arx2503_13657}, and post-hoc step localization~\cite{zhang2025which,arx2505_08638}. Replaying trajectories through temporal-logic monitors shows how missing approvals and timestamps limit runtime verification~\cite{arx2610_09793}. Our audit records build on tamper-evident logging~\cite{crosby2009efficient} and complement recent agent security controls that interpose mainly on tool execution (Section~\ref{sec:security}). Rather than logging events solely by lifecycle stage or tool call, we instrument the harness at state transfers---including compaction, memory consolidation, and rollback---so each emitted record directly supports a deterministic or model-based check and a runtime control.

%=====================================================================
\section{Limitations and Open Problems}\label{sec:limits}

Each row of Table~\ref{tab:audit} pairs a transfer failure with a check and control; while our reference middleware measures deterministic overhead, detection, and attribution on controlled fault-injection trajectories, live LLM workloads across production infrastructure and multimodal web and desktop interfaces require broader end-to-end evaluation.

Several problems remain open. Checking constraint retention needs a model call whose cost and recall on real failures are unmeasured. Web and office tasks lack a test oracle for memory writes, leaving open whether replay admission or reversible writes~\cite{su2026chronomem} work better. Write-time labelers can be attacked, commercial APIs that hide server KV state prevent direct cache-rollback checks, and audit records hold user data that require explicit retention and access policies.

%=====================================================================
\section{Conclusion}\label{sec:conclusion}

Many failures in long-horizon agents occur when the harness transfers state. Observations and memory writes admit injected content, compaction and consolidation drop constraints and provenance, and rollbacks leave KV state on the inference server. Treating each transfer as an audit point specifies what the harness should record, which checks are deterministic, and which controls those records support. The same records serve execution-trace analysis, access control, information flow control, guardrails, and recovery; authorization in particular belongs in durable state outside the prompt. Evaluating these controls on live production deployments and extending them to multimodal web and desktop agents are the next steps.

\end{document}